\documentclass[twocolumn,prl,floatfix,superscriptaddress,showpacs]{revtex4}
\usepackage{graphicx}
\usepackage{times}
\usepackage{color}
\usepackage{amsmath}
\def\D{{\rm d}}

\def\I{{\rm i \,}}
\begin{document}
\title{Chemo-mechanical instabilities in polarizable active layers}
\author{M.~H. K\"opf}
\affiliation{Department of Chemical Engineering, Technion -- Israel Institute of Technology, Haifa 32000, Israel}
\author{L.~M. Pismen}
\affiliation{Department of Chemical Engineering, Technion -- Israel Institute of Technology, Haifa 32000, Israel}
\affiliation{Minerva Center for Nonlinear Physics of Complex Systems}
\begin{abstract}
We formulate and explore a generic continuum model of a polarizable active layer with neo-Hookean elasticity and chemo-mechanical interactions. Homogeneous solutions of the model equations exhibit a stationary long-wave instability when the medium is activated by expansion, and an oscillatory short-wave instability  in the case of compressive activation. Both regimes are investigated analytically and
numerically. The long-wave instability initiates a coarsening process, which provides a possible mechanism for the establishment of permanent polarization in spherical geometry.
\end{abstract}
 \maketitle

Active media are capable to modify their mechanical properties under the influence of chemical signaling; in their turn, chemical transformations may be affected by local stress, strain and polarization. This creates a feedback loop that, on the one hand, makes rheology of the medium flexible and responsive to the environment and, on the other hand, may cause various instabilities, leading to spatial inhomogeneities, oscillations, and spontaneous motion. The primary example of an active biological medium is \emph{cytoskeleton}, which is the principal structural element of eukaryotic cells responsible for their integrity, reshaping, proliferation, and motion \cite{Howard,Lieleg2010}.

Experimental studies of living tissues give many examples of instabilities involving chemo-mechanical interactions, such as spontaneous oscillatory contraction of muscle fibers \cite{Fujita98}, traveling waves  \cite{wave04}, and intermittent localized oscillations \emph{in vivo} \cite{Martin10} that occur, in particular, during such tissue reshaping processes as dorsal closure \cite{Brunner,Gorfinkiel11} and elongation of the \emph{Drosophila} egg shell \cite{Montell10}. Inhomogeneities and patterns of non-genetic origin are inconspicuous \emph{in vivo} but are prominent in model systems, seen as clusters, swirls and interconnected bands in motility assays \cite{Bausch10}, localized motor activity spots in reconstituted actin-myosin mixtures \cite{SDS_PNAS_10}, and velocity spurts in spreading tissues \cite{silberzan10}. Mechanical impact on chemical oscillations (though without feedback reaction) has been also detected in inorganic gels \cite{KYB_SoftMatter_07}. 

Mechanisms of chemo-mechanical interactions in living cells and tissues are variegated and largely unknown but underlying general characteristics can be understood by taking as a basis well-established principles of continuous mechanics that account for the symmetries of the medium, and imposing a simple chemical feedback. The continuum approximation is particularly suitable for analytical studies of instability thresholds. Symmetry-breaking and oscillatory instabilities due to chemo-mechanical coupling have been detected in this way in the context of calcium transfer dynamics \cite{MO_JMedBio_84,OsterOdell84}, and much later with myosin motors acting as chemical variables \cite{BMC_SoftMatter_11}. Further studies included nematic polarization in the continuum theory \cite{KJJ_EPJE_05,JKP_PhysRep_07} based on de Gennes' description of nematic liquid crystals \cite{degennes} and adding a term proportional to chemical potential to account for the biological activity.  The resulting equations of polar gels are quite formidable but their simplified versions have enabled elegant description of structures and defects in nematic gels \cite{JKP_PhysRep_07}, spontaneous flow \cite{kruse11,jul-couette12}, and other phenomena in polar active media.

In this Letter, we explore an alternative case of a medium with \emph{vector} polarization. The source of polarization of cytoskeleton is directionality of actin filaments characterized by a unique direction of treadmilling and motion of myosin motors. In acto-myosin stress fibers, nematic order prevails due to periodic polarity alternation as, for example, in sarcomeric structures. On the other hand, stress fibers in motile cells have a graded polarity pattern \cite{Deguchi09} implying vector polarization.  This kind of polarization is essential for generation of \emph{active} forces that play a primary role in tissue spreading \cite{woundheal,silberzan07,salm}. Another distinctive feature of our approach is allowing for spontaneous emergence of polarization, alongside with mechanical deformation and chemical activity, whereas in earlier studies of polar active media, with few exceptions \cite{joanny09}, only the direction, rather than the absolute value of the nematic order parameter was evolving dynamically. After formulating dynamic equations applicable to a thin flat polarizable deformable layer on a solid substrate, we will determine analytically the various instabilities thresholds, and further explore numerically the evolution of emerging patterns using the neo-Hookean formulation applicable to strongly deformable soft matter \cite{Ogden}.

Dynamics of thin layers dominated by friction is Aristotelian (overdamped), with velocity rather than acceleration proportional to the force. Close to equilibrium, the force driving the evolution of any dynamic variable $\Psi$ can be derived from an appropriate energy functional ${\cal F} \{\Psi\}$, leading to a dynamic equation in the variational form $\Gamma\,\mathrm{D}_t \Psi = -\delta {\cal F}/\delta \Psi$, where D$_t=\partial_t + \mathrm{v} \cdot \nabla$ is the substantial derivative, \textbf{v} is the local velocity, and $\Gamma$ is an appropriate friction coefficient. We will write in this form basic dynamic equations for the deformation and polarization fields, and further add non-equilibrium terms accounting for the active character of the medium.  

The medium is assumed to be incompressible but it can stretch in the horizontal plane $\mathbf{x}$, while modifying the layer thickness $h$. Since the vertical deformation is small compared to the deformations in the horizontal plane and the latter, in turn, are uniform in the vertical direction, the elastic energy is expressed through the two-dimensional (2D) strain tensor. Let $\mathbf{x}, \, \mathbf{X}(\mathbf{x})$ be the dimensionless position vectors, respectively, in the reference (stress-free) and deformed states. The components of the strain tensor are $\omega_{ij} =  \partial_i X_{k}  \partial_j X_{k}-\delta_{ij}$, where $\delta_{ij}$ is the Kronecker delta; summation over repeated indices is presumed here and below.  The elastic energy is
\begin{equation}
\mathcal{F}_\mathrm{el}  =\int \mathcal{L}_\mathrm{el}  \D^2 \mathbf{x},  \quad
 {\cal L}_\mathrm{el}  =  \mu h {\cal L}_0,  \quad {\cal L}_0=\frac{1}{4}\omega_{ij}\omega_{ij}.
\label{eq:elasf}\end{equation} 
where $\mu$ is the shear modulus. 

Unlike the standard 2D formulation, the variation of $h$ should be taken into account when the dynamic equations are derived. In view of the incompressibility, $h=h_0 J^{-1}$, where $h_0$ is the unperturbed thickness and $J$ is the determinant of the 2D matrix with the components $ \partial_j X_{i}$. We write the dynamic equations in the dimensionless form
\begin{equation}
\eta \, \mathrm{D}_t X_i = \frac{1}{2} \partial_j 
  \left(\frac{\omega_{jk}}{J} \partial_k X_i  -\frac{{\cal L}_0 }{J^2}
  \epsilon_{ik} \epsilon_{jl} \partial_l X_k \right) + \frac{ f_i}{2J},
\label{eq:elas} \end{equation}
where $\mathbf{F}= (\mu/a_x)\,\mathbf{f}$ is an active force per unit volume (to be specified below), $\eta=\Gamma_\mathrm{el} a_x^2/(2\mu h_0 a_t)$, $a_x, \, a_t$ are the length and time scales, and $\epsilon_{ik}$ is the 2D antisymmetric matrix. 

Planar polarization is characterized by the 2D vector order parameter $\mathbf{p}(\mathbf{X},t)$. The polarization energy is expressed as an integral over the \emph{deformed} domain: $\mathcal{F}_\mathrm{p}  =\int  \mathcal{L}_\mathrm{p}  \D^2 \mathbf{X}$, where the Lagrangian ${\cal L}_p$ should be expressed through 
rotationally invariant combinations of $\mathbf{p}$ and the 2D gradient operator $\nabla$ with respect to the coordinates $X_i$. Retaining terms with the added orders in $\mathbf{p}$ and $\nabla$ not exceeding four, we write
\begin{eqnarray}
{\cal L}_\mathbf{p} &=&  \frac{\kappa h}{2 a_x^2} \left[  \alpha^2
 |\mathbf{p}|^2  \left(1+ \frac{ \kappa_0}{2} |\mathbf{p}|^2  \right)
+ 2 \kappa_{1}  \nabla_i p_i 
 \right.  \nonumber \\
&+& \left.  ( \nabla_i p_i)^2  + \kappa_{2}( \epsilon_{ik} \nabla_i  p_k)^2
 + \kappa_{3} |\mathbf{p}|^2  \nabla_i p_i \right].
\label{2ff2}
\end{eqnarray}
The second term is irrelevant, unless the coefficient $ \kappa_{1}$ couples the order parameter with another field. In an active medium, this might be the concentration $c$ of some chemical species. We choose $\kappa_0^{-1/2}$ as the polarization scale and set $\kappa_{1}=- \kappa_0^{-1/2}c$, where $c(\mathbf{X},t)$ is interpreted as a scaled deviation from a reference concentration $c_0$. Setting also $\kappa_3=0$, we write the dynamic equation generated by the Lagrangian \eqref{2ff2} in the dimensionless form 
\begin{eqnarray}
 \gamma  \,\mathrm{D}_t p_i &=& \nabla_i (J^{-1}   \nabla_j p_j)  
 + \kappa_{2}  \nabla_j\left[  J^{-1}\left( \nabla_j p_i - \nabla_i p_j\right) \right]
 \nonumber \\
& -&  \nabla_i (J^{-1} c) - \alpha^2 J^{-1}\left(1 + |\mathbf{p}|^2  \right)p_i,
\label{eq:pevol}  
\end{eqnarray}
where $\gamma=\Gamma_\mathbf{p} a_x^2/( \kappa  h_0 a_t)$. Generally, one has to include the variation of the polarization energy into the deformation equations; indeed, a polarized medium may deform to releave polarization gradients. We will avoid, however, this complication, assuming $\kappa \ll \mu$. The coupling between polarization and deformation will be established instead by the active force in Eq.~\eqref{eq:elas} directed along the polarization direction, $f_i = 2q p_i$.

The feedback loop is closed by making concentration of the active species strain-dependent. We write the equation of the concentration field taking into account that, due to incompressibility, the concentration is not affected by deformation; the diffusional flux is, however, proportional to the local thickness. Allowing for the production rate proportional to the local strain and linear decay with the constant identified with the inverse time scale $1/a_t$, we write the concentration equation in the dimensionless form 
\begin{equation}
\mathrm{D}_t c  = J \nabla_i (J^{-1}\nabla_i c) - c + \beta (J-1). \label{eq:cevol}
\end{equation}
where we have chosen as the length scale the diffusion length $a_x=\sqrt{D
a_t}$, $D$ being the diffusivity.

The above equation system becomes extremely convoluted if applied in a fixed coordinate frame where, besides computing the substantial derivative that includes the velocity $\mathbf{v}=\partial_t \mathbf{X}$, one would need to transform the $\nabla$ operator in Eqs.~\eqref{eq:pevol}, \eqref{eq:cevol} to account for deformations. These complications do not affect, however, linear stability analysis, and will be avoided at large deformations by implementing a Lagrangean numerical procedure.

\begin{figure}
\centering
\includegraphics{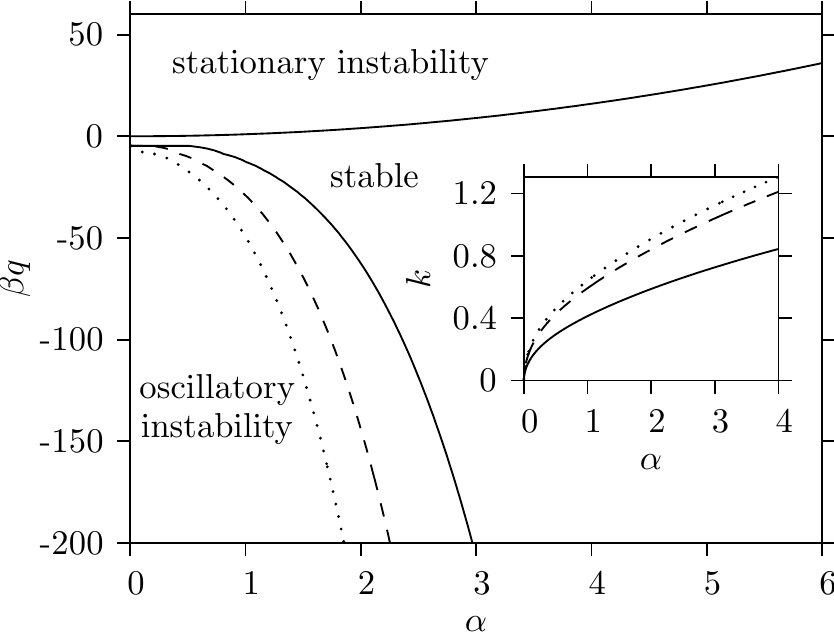}
\caption{\label{fig:stabdiag}Stability diagram in the parametric plane
 $\alpha, \:\beta q$. The oscillatory instability limits are shown for $\gamma=5/4$ and $\eta=1/3$ (solid line), $\eta=5/3$ (dashes), and $\eta=10/3$ (dots). Inset: the critical wavelength $k$ at the oscillatory bifurcation lines parameterized by $\alpha$.}
\end{figure}

We explore now stability of the homogeneous solution $\mathbf{p}=0$, $c=0$, $\mathbf{X}=\mathbf{x}$ to infinitesimal perturbations. 
The linearized system determining the bifurcation threshold can be
expressed through the 2D strain $\psi = J-1= \nabla_i (X_i-x_i)$ and splay $\phi=\nabla_i p_i$ by taking the divergence of linearized Eqs.~\eqref{eq:elas}, \eqref{eq:pevol}: 
\begin{eqnarray}
\eta \,\partial_t \psi &=&  \nabla^2\psi+q \phi,  \label{eq:psievollin} \\
\gamma \,\partial_t \phi &=&  \nabla^2 \phi  -\alpha^2 \phi -   \nabla^2 c, 
\label{eq:phievollin} \\
\partial_t c  &=& \nabla^2 c - c + \beta\psi.
 \label{eq:cevollin}
\end{eqnarray}
A nontrivial solution of Eqs.~(\ref{eq:psievollin}) -- (\ref{eq:cevollin}) is
sought for as a waveform $\exp(\lambda t +\I \mathbf{k}\cdot \mathbf{x})$ with
a wave vector $\mathbf{k}$. The eigenvalues $\lambda$ determining stability of
the uniform state $c=\psi = \phi =0$ are computed by setting to zero the determinant of the Fourier transformed linearized system. The bifurcation thresholds are computed in a standard way by requiring 
$\sup_{|\mathbf{k}|}(\Re \lambda )=0$. 

In the case of extentional activation ($\beta>0$), one can show that the branch
of real eigenvalues first reaches zero in the long-wave mode $\mathbf{k}=0$ at
$\beta q=\alpha^2$.  In the case of compressive activation  ($\beta<0$), a
\emph{wave} instability at finite $|\mathbf{k}|$ and $\Im \lambda \neq 0$  is
observed. The analytic formulas for the critical wavenumber $k_\mathrm{c}$ and
the critical value $\beta_\mathrm{c}$ are too cumbersome to be of practical use
and will not be listed here. The minimal gap width between the two instability limits equals to $\gamma+\eta$. The bifurcation diagram in the parametric plane
$\alpha, \, \beta q$ is shown in Fig.~\ref{fig:stabdiag}.

We explore dynamics beyond the bifurcation thresholds for both types of
activation by direct numerical simulation, starting from randomly perturbed
homogeneous initial conditions. We discretize Eqs.~\eqref{eq:elas},
\eqref{eq:pevol}, and \eqref{eq:cevol} using a flexible grid with nodes
following the physical displacement of the medium. In the spirit of
mesh-free computations \cite{OIZT_JNME_96}, the spatial derivatives are 
calculated based on local approximation of the fields by low-order polynomials.
The resulting set of ordinary differential equations is then solved using
an embedded adaptive Runge-Kutta-Fehlberg scheme of the order 4 and 5 \cite{NRC}. The parameters are set to $q=\sqrt{5}/100,\alpha=\sqrt{5},\kappa_2=1, \eta=5/3, \:\gamma=5/4$.

\begin{figure}[t]
\centering
\includegraphics[width=.235\textwidth]{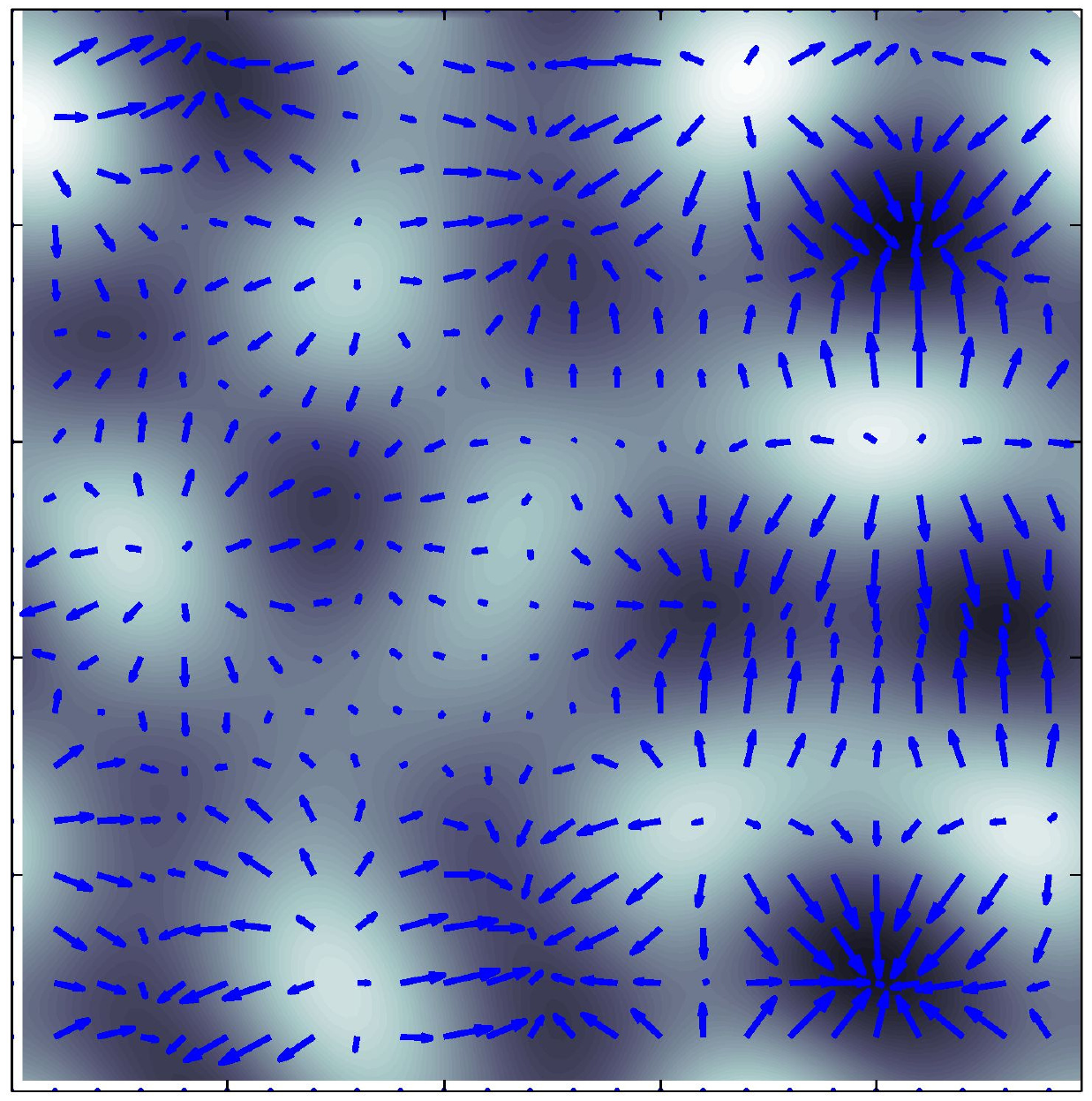}
\includegraphics[width=.235\textwidth]{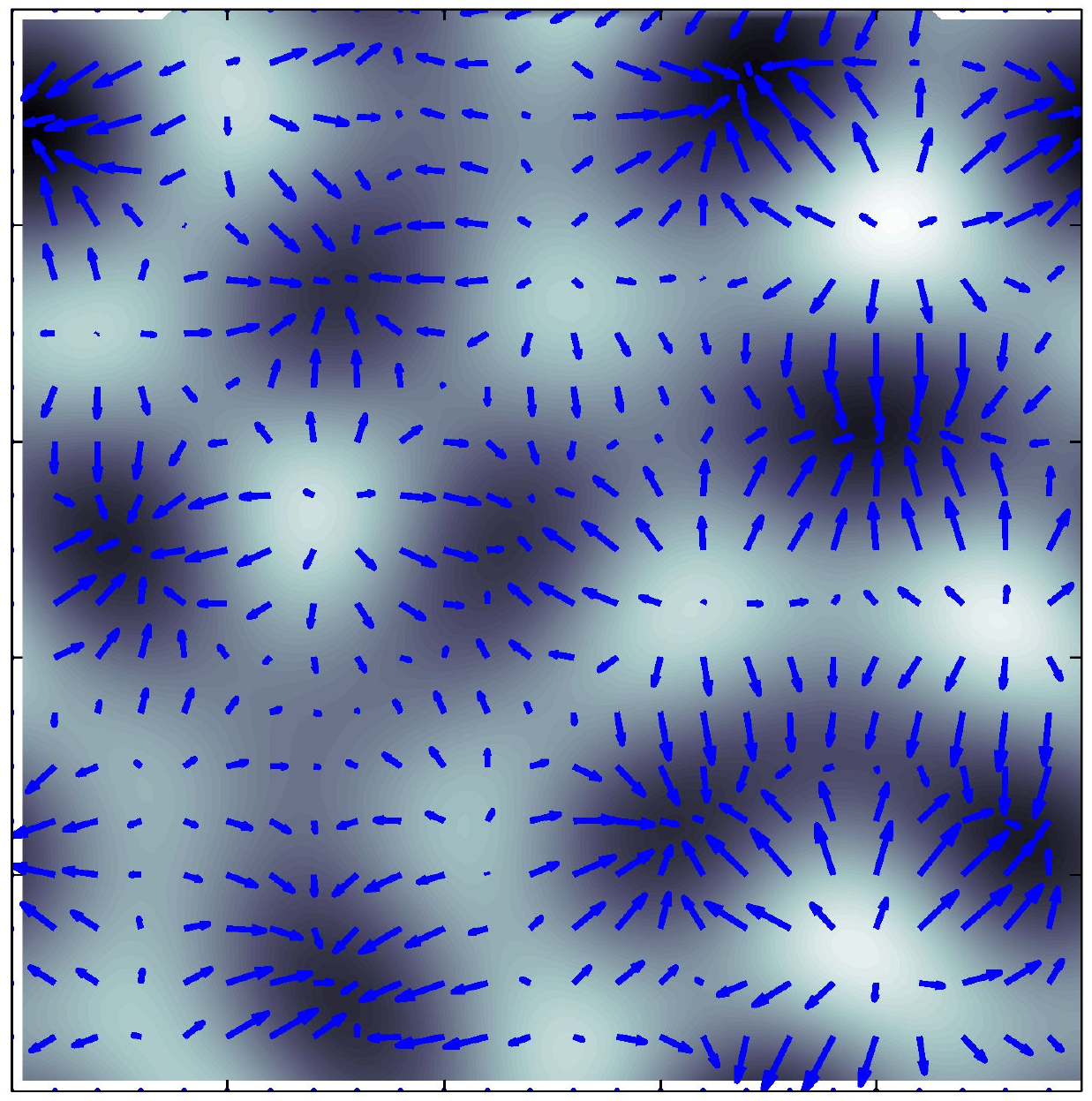}
\caption{\label{fig:osci}Snapshots from a simulation in the
oscillatory unstable parametric region ($\beta q=-210$, $(\beta q)_c\approx-196$). The time difference between the two images equals half the oscillation period.  The
simulation domain is a double-periodic $10\sqrt{5}\times10\sqrt{5}$ square.
Lighter shades correspond to higher concentration levels; scaled arrows indicate the
polarization $\mathbf{p}$.}
\end{figure}

\begin{figure}
\centering
\includegraphics[width=.42\textwidth]{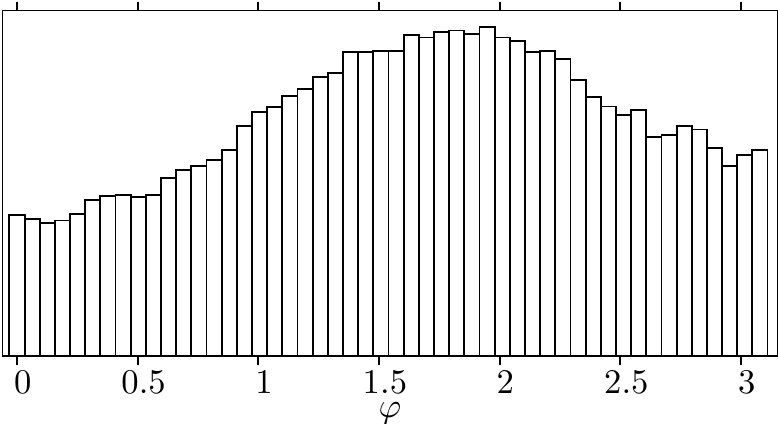}
\caption{\label{fig:osciangles} Histogram of the angle $\varphi$ between the directions of the elastic and active force.}
\end{figure}

In the oscillatory regime, we observe a wave pattern with the active force (directed from high to low $c$ regions) following the rotating concentration gradient between the domains oscillating in antiphase, as seen in Fig.~\ref{fig:osci}. The directions of the elastic and active force are, however, disaligned due to delay, so that the angle between the two is widely distributed with a shallow maximum about $\pi/2$ (Fig.~\ref{fig:osciangles}).

\begin{figure}
\centering
\includegraphics[width=.235\textwidth]{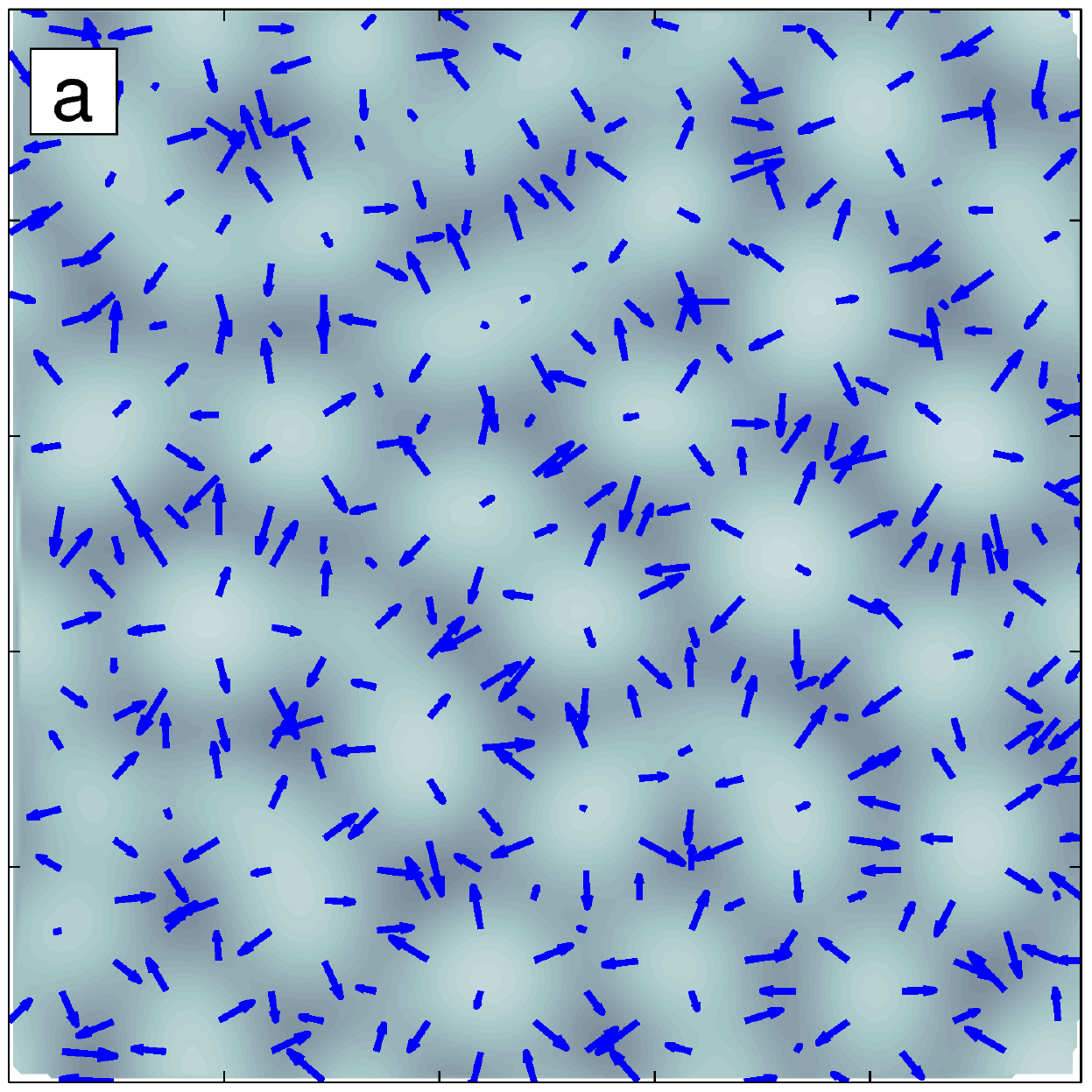}
\includegraphics[width=.235\textwidth]{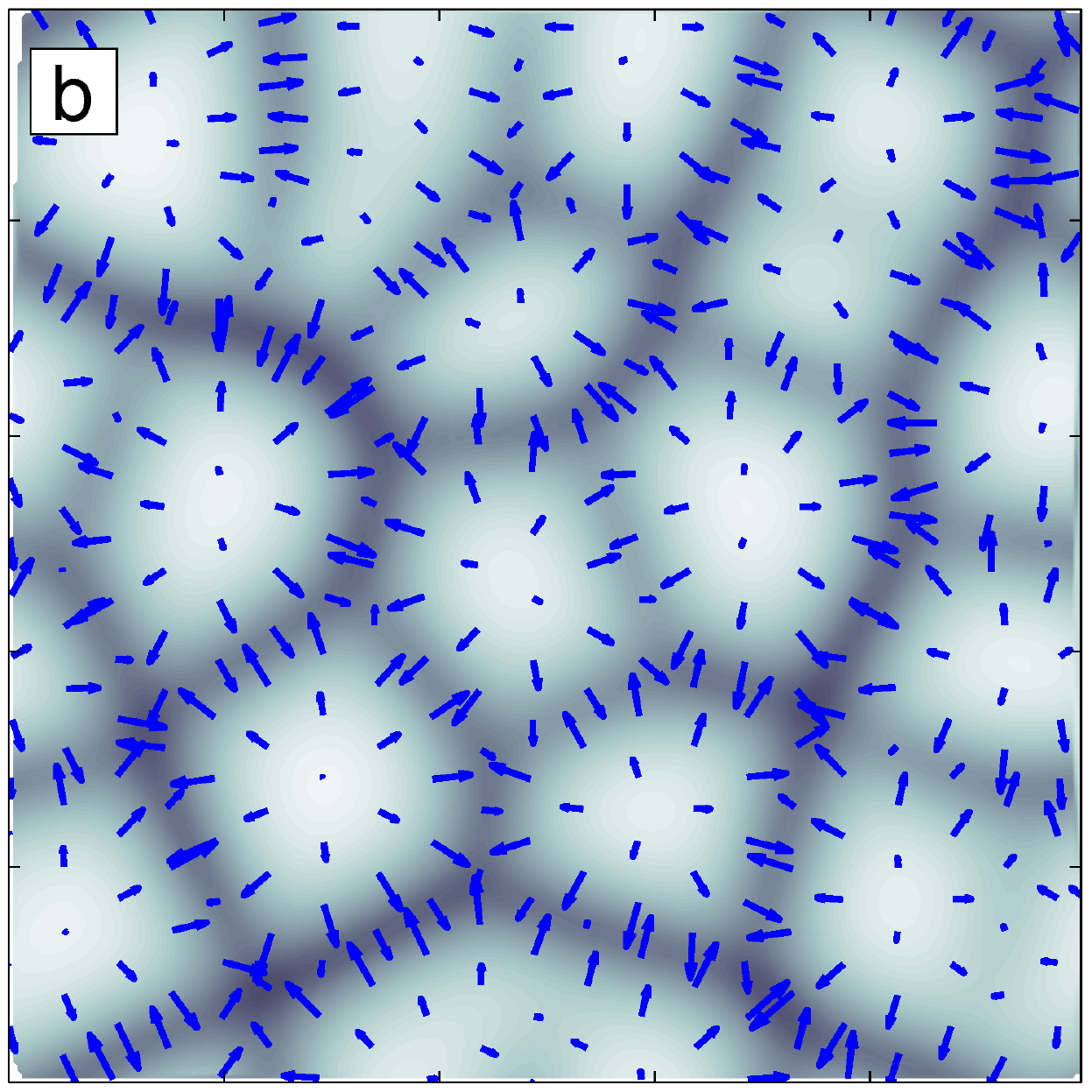}
\includegraphics[width=.235\textwidth]{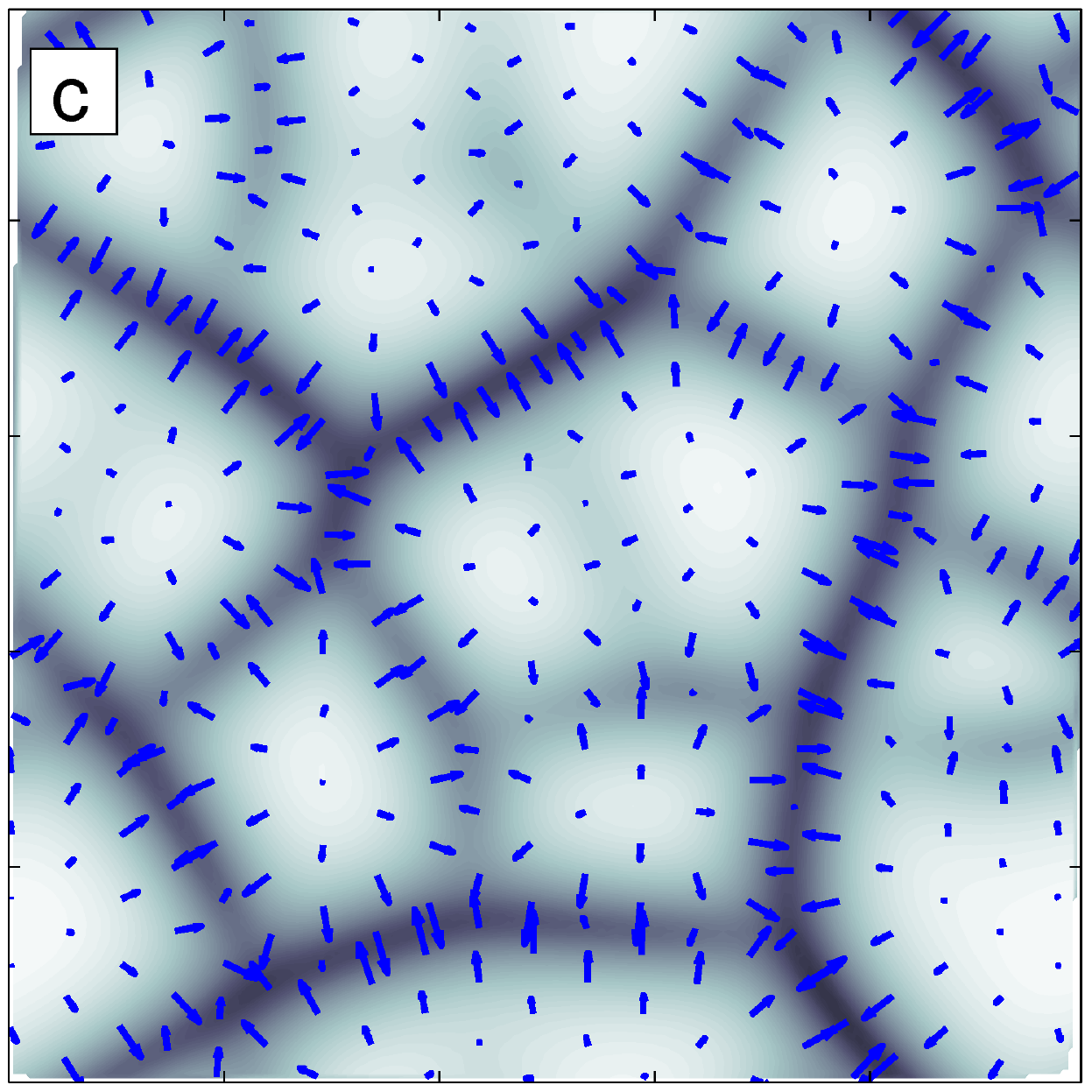}
\includegraphics[width=.235\textwidth]{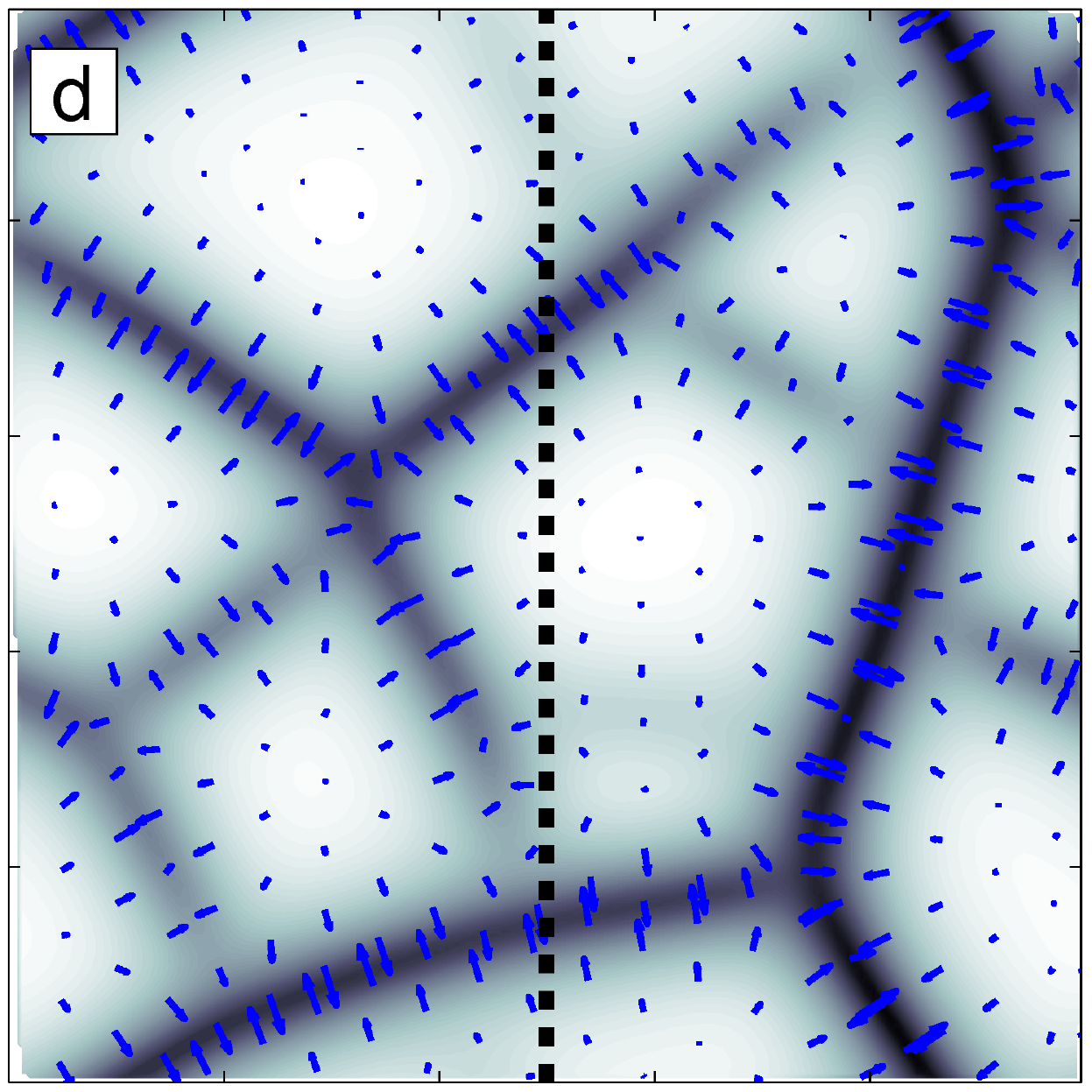}
\caption{\label{fig:coarsening} Domain coarsening: snapshots at $t=4\times10^3$ (a), $2\times10^4$ (b), $3\times10^4$ (c), $5\times10^4$ (d), simulation on the double-periodic
$100\sqrt{5}\times100\sqrt{5}$ square domain for $\beta q=6$, $(\beta q)_c=5$.}
\end{figure}
\begin{figure}
\centering
\includegraphics[width=.45\textwidth]{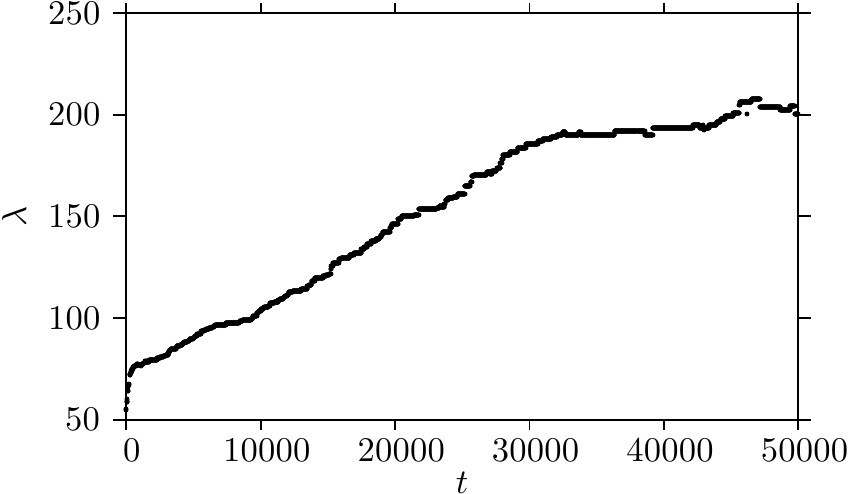}
\caption{\label{fig:lambda}Time evolution of the wavelength of the dominant
Fourier mode during coarsening (the
average of $30$ simulations on the double-periodic $100\sqrt{5}\times100\sqrt{5}$ square domain for $\beta q=6$).}
\end{figure}
In the case of extensional activation, the behavior is qualitatively
different. Initially, a short-wave pattern corresponding to a fastest-growing mode emerges from the initial noise. Then, a slow coarsening process ensues, during which larger domains grow while smaller ones either merge or vanish, so that the dominant wavelength grows, as seen in Fig.~\ref{fig:coarsening}. 
The wavelength corresponding to the maximum of the Fourier spectrum is plotted
against time in Fig.~\ref{fig:lambda}. Due to the finite size of the simulation
domain, the plot for each individual run has a number of discontinuous steps
but it approaches a monotonically growing curve when averaged over $30$ runs. 

\begin{figure}
\centering
\includegraphics[width=.4\textwidth]{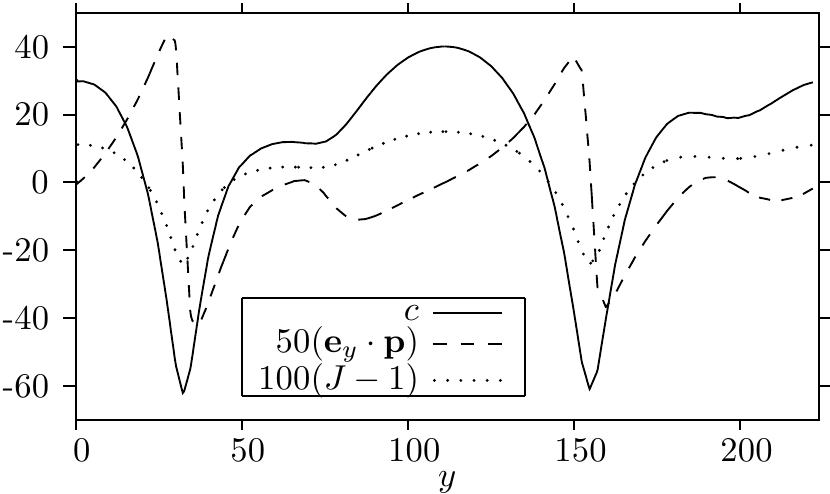}
\caption{\label{fig:rift}Cross section through the dashed line in Fig.~\ref{fig:coarsening} (d). The polarization component along the line $\mathbf{p} \cdot \mathbf{e}_y$ and deformation $J-1$ are scaled to show all three fields in the same plot.}
\end{figure}

Unlike the oscillatory case, the active and elastic forces are almost perfectly anti-aligned: the mutual angle deviates from $\pi$ by less than $3.5^\circ$ with $98\%$ probability.
As the active force is directed against the concentration gradient, the low-$c$ 
regions are squeezed into narrow stripes separating high-$c$ domains. In turn,   the chemical production is inhibited within these stripes due to strong compression, as the deformation acts as the source of $c$ in Eq.~\eqref{eq:cevol}. The resulting correlation between $J$ and $c$ is seen in Fig.~\ref{fig:rift} that shows the cross-section profiles at a late stage of coarsening.
\begin{figure}
\centering
\includegraphics[width=.42\textwidth]{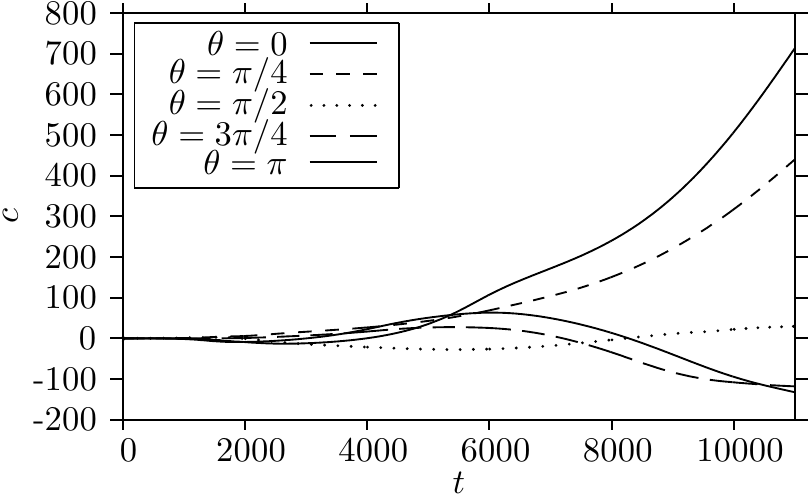}
\caption{Time evolution of the concentration $c$ at different polar angles $\theta$ on a sphere of radius $10\sqrt{5}$.\label{fig:spherelines}}
\end{figure}

The coarsening process provides a natural mechanism for polarity establishment in
cortical tissues, as is observed, for example, in the \emph{C.~elegans} embryo \cite{c-eleg}. When Eqs.~\eqref{eq:elas}, \eqref{eq:pevol}, and \eqref{eq:cevol} are solved on a spherical surface, the coarsening consolidates two domains, whereafter the smaller one is compressed into a low-$c$ spot. This is illustrated in Fig.~\ref{fig:spherelines}, where the evolution of $c$ is plotted for five
different polar angles $\theta$, where $\theta=0$ is defined by the location of
the high-concentration pole. As the polarization is aligned with the concentration gradient, the active force is directed from one pole to the other. The established polarity axis is \emph{permanent}, unlike Ref.~\cite{Grill11} where advection serves as a mere trigger requiring additional means for preserving the polarization.

\acknowledgments
{This work has been supported by the Human Frontier Science Program (Grant RGP0052/2009-C). }



\begin{thebibliography}{30}
\expandafter\ifx\csname natexlab\endcsname\relax\def\natexlab#1{#1}\fi
\expandafter\ifx\csname bibnamefont\endcsname\relax
  \def\bibnamefont#1{#1}\fi
\expandafter\ifx\csname bibfnamefont\endcsname\relax
  \def\bibfnamefont#1{#1}\fi
\expandafter\ifx\csname citenamefont\endcsname\relax
  \def\citenamefont#1{#1}\fi
\expandafter\ifx\csname url\endcsname\relax
  \def\url#1{\texttt{#1}}\fi
\expandafter\ifx\csname urlprefix\endcsname\relax\def\urlprefix{URL }\fi
\providecommand{\bibinfo}[2]{#2}
\providecommand{\eprint}[2][]{\url{#2}}

\bibitem[{\citenamefont{Howard}(2001)}]{Howard}
\bibinfo{author}{\bibfnamefont{J.}~\bibnamefont{Howard}},
  \emph{\bibinfo{title}{Mechanics of Motor Proteins and the Cytoskeleton}}
  (\bibinfo{publisher}{Sinauer Associates}, \bibinfo{address}{Sunderland,
  Mass.}, \bibinfo{year}{2001}).

\bibitem[{\citenamefont{Lieleg et~al.}(2010)\citenamefont{Lieleg, Claessens,
  and Bausch}}]{Lieleg2010}
\bibinfo{author}{\bibfnamefont{O.}~\bibnamefont{Lieleg}},
  \bibinfo{author}{\bibfnamefont{M.~M. A.~E.} \bibnamefont{Claessens}},
  \bibnamefont{and} \bibinfo{author}{\bibfnamefont{A.~R.}
  \bibnamefont{Bausch}}, \bibinfo{journal}{Soft Matter}
  \textbf{\bibinfo{volume}{6}}, \bibinfo{pages}{218} (\bibinfo{year}{2010}).

\bibitem[{\citenamefont{Fujita and Ishiwata}(1998)}]{Fujita98}
\bibinfo{author}{\bibfnamefont{H.}~\bibnamefont{Fujita}} \bibnamefont{and}
  \bibinfo{author}{\bibfnamefont{S.}~\bibnamefont{Ishiwata}},
  \bibinfo{journal}{Biophys. J.} \textbf{\bibinfo{volume}{75}},
  \bibinfo{pages}{1439} (\bibinfo{year}{1998}).

\bibitem[{\citenamefont{Gerisch et~al.}(2004)\citenamefont{Gerisch,
  Bretschneider, and Muller-Taubenberger}}]{wave04}
\bibinfo{author}{\bibfnamefont{G.}~\bibnamefont{Gerisch}},
  \bibinfo{author}{\bibfnamefont{T.}~\bibnamefont{Bretschneider}},
  \bibnamefont{and}
  \bibinfo{author}{\bibfnamefont{A.}~\bibnamefont{Muller-Taubenberger}},
  \bibinfo{journal}{Biophys. J.} \textbf{\bibinfo{volume}{87}},
  \bibinfo{pages}{3493} (\bibinfo{year}{2004}).

\bibitem[{\citenamefont{Martin}(2010)}]{Martin10}
\bibinfo{author}{\bibfnamefont{A.~C.} \bibnamefont{Martin}},
  \bibinfo{journal}{Developmental Biol.} \textbf{\bibinfo{volume}{341}},
  \bibinfo{pages}{114} (\bibinfo{year}{2010}).

\bibitem[{\citenamefont{Solon et~al.}(2009)\citenamefont{Solon, Kaya-\c{C}opur,
  Colombelli, and Brunner}}]{Brunner}
\bibinfo{author}{\bibfnamefont{J.}~\bibnamefont{Solon}},
  \bibinfo{author}{\bibfnamefont{A.}~\bibnamefont{Kaya-\c{C}opur}},
  \bibinfo{author}{\bibfnamefont{J.}~\bibnamefont{Colombelli}},
  \bibnamefont{and} \bibinfo{author}{\bibfnamefont{D.}~\bibnamefont{Brunner}},
  \bibinfo{journal}{Cell} \textbf{\bibinfo{volume}{137}}, \bibinfo{pages}{1331}
  (\bibinfo{year}{2009}).

\bibitem[{\citenamefont{Gorfinkiel et~al.}(2011)\citenamefont{Gorfinkiel,
  Schamberg, and Blanchard}}]{Gorfinkiel11}
\bibinfo{author}{\bibfnamefont{N.}~\bibnamefont{Gorfinkiel}},
  \bibinfo{author}{\bibfnamefont{S.}~\bibnamefont{Schamberg}},
  \bibnamefont{and} \bibinfo{author}{\bibfnamefont{G.~B.}
  \bibnamefont{Blanchard}}, \bibinfo{journal}{Genesis}
  \textbf{\bibinfo{volume}{49}}, \bibinfo{pages}{522} (\bibinfo{year}{2011}).

\bibitem[{\citenamefont{He et~al.}(2010)\citenamefont{He, Wang, Tang, and
  Montell}}]{Montell10}
\bibinfo{author}{\bibfnamefont{L.}~\bibnamefont{He}},
  \bibinfo{author}{\bibfnamefont{X.}~\bibnamefont{Wang}},
  \bibinfo{author}{\bibfnamefont{H.~L.} \bibnamefont{Tang}}, \bibnamefont{and}
  \bibinfo{author}{\bibfnamefont{D.~J.} \bibnamefont{Montell}},
  \bibinfo{journal}{Nature Cell Biol.} \textbf{\bibinfo{volume}{12}},
  \bibinfo{pages}{1133} (\bibinfo{year}{2010}).

\bibitem[{\citenamefont{Schaller et~al.}(2010)\citenamefont{Schaller, Weber,
  Semmrich, Frey, and Bausch}}]{Bausch10}
\bibinfo{author}{\bibfnamefont{V.}~\bibnamefont{Schaller}},
  \bibinfo{author}{\bibfnamefont{C.}~\bibnamefont{Weber}},
  \bibinfo{author}{\bibfnamefont{C.}~\bibnamefont{Semmrich}},
  \bibinfo{author}{\bibfnamefont{E.}~\bibnamefont{Frey}}, \bibnamefont{and}
  \bibinfo{author}{\bibfnamefont{A.~R.} \bibnamefont{Bausch}},
  \bibinfo{journal}{Nature} \textbf{\bibinfo{volume}{467}}, \bibinfo{pages}{73}
  (\bibinfo{year}{2010}).

\bibitem[{\citenamefont{Silva et~al.}(2011)\citenamefont{Silva, Depken,
  Stuhrmann, Korsten, MacKintosh, and Koenderink}}]{SDS_PNAS_10}
\bibinfo{author}{\bibfnamefont{M.~S.~e.} \bibnamefont{Silva}},
  \bibinfo{author}{\bibfnamefont{M.}~\bibnamefont{Depken}},
  \bibinfo{author}{\bibfnamefont{B.}~\bibnamefont{Stuhrmann}},
  \bibinfo{author}{\bibfnamefont{M.}~\bibnamefont{Korsten}},
  \bibinfo{author}{\bibfnamefont{F.~C.} \bibnamefont{MacKintosh}},
  \bibnamefont{and} \bibinfo{author}{\bibfnamefont{G.~H.}
  \bibnamefont{Koenderink}}, \bibinfo{journal}{Proc. Natl Acad. Sci. U.S.A.}
  \textbf{\bibinfo{volume}{108}}, \bibinfo{pages}{9408} (\bibinfo{year}{2011}).

\bibitem[{\citenamefont{Petitjean et~al.}(2010)\citenamefont{Petitjean, Reffay,
  Grasland-Mongrain, Poujade, Ladoux, Buguin, and Silberzan}}]{silberzan10}
\bibinfo{author}{\bibfnamefont{L.}~\bibnamefont{Petitjean}},
  \bibinfo{author}{\bibfnamefont{M.}~\bibnamefont{Reffay}},
  \bibinfo{author}{\bibfnamefont{E.}~\bibnamefont{Grasland-Mongrain}},
  \bibinfo{author}{\bibfnamefont{M.}~\bibnamefont{Poujade}},
  \bibinfo{author}{\bibfnamefont{B.}~\bibnamefont{Ladoux}},
  \bibinfo{author}{\bibfnamefont{A.}~\bibnamefont{Buguin}}, \bibnamefont{and}
  \bibinfo{author}{\bibfnamefont{P.}~\bibnamefont{Silberzan}},
  \bibinfo{journal}{Biophys J} \textbf{\bibinfo{volume}{98}},
  \bibinfo{pages}{1790} (\bibinfo{year}{2010}).

\bibitem[{\citenamefont{Kuksenok et~al.}(2007)\citenamefont{Kuksenok, Yashin,
  and Balazs}}]{KYB_SoftMatter_07}
\bibinfo{author}{\bibfnamefont{O.}~\bibnamefont{Kuksenok}},
  \bibinfo{author}{\bibfnamefont{V.~V.} \bibnamefont{Yashin}},
  \bibnamefont{and} \bibinfo{author}{\bibfnamefont{A.~C.}
  \bibnamefont{Balazs}}, \bibinfo{journal}{Soft Matter}
  \textbf{\bibinfo{volume}{3}}, \bibinfo{pages}{1138} (\bibinfo{year}{2007}).

\bibitem[{\citenamefont{Murray and Oster}(1984)}]{MO_JMedBio_84}
\bibinfo{author}{\bibfnamefont{J.~D.} \bibnamefont{Murray}} \bibnamefont{and}
  \bibinfo{author}{\bibfnamefont{G.~F.} \bibnamefont{Oster}},
  \bibinfo{journal}{IMA J. Math. Appl. Math. Med. Biol.}
  \textbf{\bibinfo{volume}{1}}, \bibinfo{pages}{51} (\bibinfo{year}{1984}).

\bibitem[{\citenamefont{Oster and Odell}(1984)}]{OsterOdell84}
\bibinfo{author}{\bibfnamefont{G.}~\bibnamefont{Oster}} \bibnamefont{and}
  \bibinfo{author}{\bibfnamefont{G.}~\bibnamefont{Odell}},
  \bibinfo{journal}{Physica D} \textbf{\bibinfo{volume}{12}},
  \bibinfo{pages}{333} (\bibinfo{year}{1984}).

\bibitem[{\citenamefont{Banerjee and Marchetti}(2011)}]{BMC_SoftMatter_11}
\bibinfo{author}{\bibfnamefont{S.}~\bibnamefont{Banerjee}} \bibnamefont{and}
  \bibinfo{author}{\bibfnamefont{M.~C.} \bibnamefont{Marchetti}},
  \bibinfo{journal}{Soft Matter} \textbf{\bibinfo{volume}{7}},
  \bibinfo{pages}{463} (\bibinfo{year}{2011}).

\bibitem[{\citenamefont{Kruse et~al.}(2005)\citenamefont{Kruse, Joanny,
  J\"ulicher, Prost, and Sekimoto}}]{KJJ_EPJE_05}
\bibinfo{author}{\bibfnamefont{K.}~\bibnamefont{Kruse}},
  \bibinfo{author}{\bibfnamefont{J.~F.} \bibnamefont{Joanny}},
  \bibinfo{author}{\bibfnamefont{F.}~\bibnamefont{J\"ulicher}},
  \bibinfo{author}{\bibfnamefont{J.}~\bibnamefont{Prost}}, \bibnamefont{and}
  \bibinfo{author}{\bibfnamefont{K.}~\bibnamefont{Sekimoto}},
  \bibinfo{journal}{Eur. Phys. J. E} \textbf{\bibinfo{volume}{16}},
  \bibinfo{pages}{5} (\bibinfo{year}{2005}).

\bibitem[{\citenamefont{J\"ulicher et~al.}(2007)\citenamefont{J\"ulicher,
  Kruse, Prost, and Joanny}}]{JKP_PhysRep_07}
\bibinfo{author}{\bibfnamefont{F.}~\bibnamefont{J\"ulicher}},
  \bibinfo{author}{\bibfnamefont{K.}~\bibnamefont{Kruse}},
  \bibinfo{author}{\bibfnamefont{J.}~\bibnamefont{Prost}}, \bibnamefont{and}
  \bibinfo{author}{\bibfnamefont{J.-F.} \bibnamefont{Joanny}},
  \bibinfo{journal}{Phys. Rep.} \textbf{\bibinfo{volume}{449}},
  \bibinfo{pages}{3} (\bibinfo{year}{2007}).

\bibitem[{\citenamefont{de~Gennes and Prost}(1993)}]{degennes}
\bibinfo{author}{\bibfnamefont{P.}~\bibnamefont{de~Gennes}} \bibnamefont{and}
  \bibinfo{author}{\bibfnamefont{J.}~\bibnamefont{Prost}},
  \emph{\bibinfo{title}{The Physics of Liquid Crystals}}
  (\bibinfo{publisher}{Oxford University Press}, \bibinfo{address}{Oxford,
  U.K.}, \bibinfo{year}{1993}).

\bibitem[{\citenamefont{Doubrovinski and Kruse}(2011)}]{kruse11}
\bibinfo{author}{\bibfnamefont{K.}~\bibnamefont{Doubrovinski}}
  \bibnamefont{and} \bibinfo{author}{\bibfnamefont{K.}~\bibnamefont{Kruse}},
  \bibinfo{journal}{Phys. Rev. Lett.} \textbf{\bibinfo{volume}{107}},
  \bibinfo{pages}{258103} (\bibinfo{year}{2011}).

\bibitem[{\citenamefont{F\"urthauer et~al.}(2012)\citenamefont{F\"urthauer,
  Neef, Grill, Kruse, and J\"ulicher}}]{jul-couette12}
\bibinfo{author}{\bibfnamefont{S.}~\bibnamefont{F\"urthauer}},
  \bibinfo{author}{\bibfnamefont{M.}~\bibnamefont{Neef}},
  \bibinfo{author}{\bibfnamefont{S.~W.} \bibnamefont{Grill}},
  \bibinfo{author}{\bibfnamefont{K.}~\bibnamefont{Kruse}}, \bibnamefont{and}
  \bibinfo{author}{\bibfnamefont{F.}~\bibnamefont{J\"ulicher}},
  \bibinfo{journal}{New J. Phys.} \textbf{\bibinfo{volume}{14}},
  \bibinfo{pages}{023001} (\bibinfo{year}{2012}).

\bibitem[{\citenamefont{Deguchi and Sato}(2009)}]{Deguchi09}
\bibinfo{author}{\bibfnamefont{S.}~\bibnamefont{Deguchi}} \bibnamefont{and}
  \bibinfo{author}{\bibfnamefont{M.}~\bibnamefont{Sato}},
  \bibinfo{journal}{Biorheology} \textbf{\bibinfo{volume}{46}},
  \bibinfo{pages}{93} (\bibinfo{year}{2009}).

\bibitem[{\citenamefont{Nikolic et~al.}(2006)\citenamefont{Nikolic, Boettiger,
  Bar-Sagi, Carbeck, and Shvartsman}}]{woundheal}
\bibinfo{author}{\bibfnamefont{D.~L.} \bibnamefont{Nikolic}},
  \bibinfo{author}{\bibfnamefont{A.~N.} \bibnamefont{Boettiger}},
  \bibinfo{author}{\bibfnamefont{D.}~\bibnamefont{Bar-Sagi}},
  \bibinfo{author}{\bibfnamefont{J.~D.} \bibnamefont{Carbeck}},
  \bibnamefont{and} \bibinfo{author}{\bibfnamefont{S.~Y.}
  \bibnamefont{Shvartsman}}, \textbf{\bibinfo{volume}{291}},
  \bibinfo{pages}{C68} (\bibinfo{year}{2006}).

\bibitem[{\citenamefont{Poujade et~al.}(2007)\citenamefont{Poujade,
  Grasland-Mongrain, Hertzog, Jouanneau, Chavrier, Ladoux, Buguin, and
  Silberzan}}]{silberzan07}
\bibinfo{author}{\bibfnamefont{M.}~\bibnamefont{Poujade}},
  \bibinfo{author}{\bibfnamefont{E.}~\bibnamefont{Grasland-Mongrain}},
  \bibinfo{author}{\bibfnamefont{A.}~\bibnamefont{Hertzog}},
  \bibinfo{author}{\bibfnamefont{J.}~\bibnamefont{Jouanneau}},
  \bibinfo{author}{\bibfnamefont{P.}~\bibnamefont{Chavrier}},
  \bibinfo{author}{\bibfnamefont{B.}~\bibnamefont{Ladoux}},
  \bibinfo{author}{\bibfnamefont{A.}~\bibnamefont{Buguin}}, \bibnamefont{and}
  \bibinfo{author}{\bibfnamefont{P.}~\bibnamefont{Silberzan}},
  \bibinfo{journal}{Proc. Nat. Acad. Sci. USA} \textbf{\bibinfo{volume}{104}},
  \bibinfo{pages}{15988} (\bibinfo{year}{2007}).

\bibitem[{\citenamefont{Salm and Pismen}(2012)}]{salm}
\bibinfo{author}{\bibfnamefont{M.}~\bibnamefont{Salm}} \bibnamefont{and}
  \bibinfo{author}{\bibfnamefont{L.~M.} \bibnamefont{Pismen}},
  \bibinfo{journal}{Phys. Biol.} \textbf{\bibinfo{volume}{9}}
  (\bibinfo{year}{2012}).

\bibitem[{\citenamefont{Salbreux et~al.}(2009)\citenamefont{Salbreux, Prost,
  and Joanny}}]{joanny09}
\bibinfo{author}{\bibfnamefont{G.}~\bibnamefont{Salbreux}},
  \bibinfo{author}{\bibfnamefont{J.}~\bibnamefont{Prost}}, \bibnamefont{and}
  \bibinfo{author}{\bibfnamefont{J.~F.} \bibnamefont{Joanny}},
  \bibinfo{journal}{Phys. Rev. Lett.} \textbf{\bibinfo{volume}{103}},
  \bibinfo{eid}{058102} (\bibinfo{year}{2009}).

\bibitem[{\citenamefont{Ogden}(2001)}]{Ogden}
\bibinfo{author}{\bibfnamefont{R.~W.} \bibnamefont{Ogden}}, in
  \emph{\bibinfo{booktitle}{Nonlinear Elasticity: Theory and Applications}},
  edited by \bibinfo{editor}{\bibfnamefont{Y.~B.} \bibnamefont{Fu}}
  \bibnamefont{and} \bibinfo{editor}{\bibfnamefont{R.~W.} \bibnamefont{Ogden}}
  (\bibinfo{publisher}{Cambridge University Press},
  \bibinfo{address}{Cambridge, U.K.}, \bibinfo{year}{2001}), pp.
  \bibinfo{pages}{1--58}.

\bibitem[{\citenamefont{Onate et~al.}(1996)\citenamefont{Onate, Idelsohn,
  Zienkiewicz, and Taylor}}]{OIZT_JNME_96}
\bibinfo{author}{\bibfnamefont{E.}~\bibnamefont{Onate}},
  \bibinfo{author}{\bibfnamefont{S.}~\bibnamefont{Idelsohn}},
  \bibinfo{author}{\bibfnamefont{O.~C.} \bibnamefont{Zienkiewicz}},
  \bibnamefont{and} \bibinfo{author}{\bibfnamefont{R.~L.}
  \bibnamefont{Taylor}}, \bibinfo{journal}{Int. J. Numer. Meth. Engng.}
  \textbf{\bibinfo{volume}{39}}, \bibinfo{pages}{3839} (\bibinfo{year}{1996}).

\bibitem[{\citenamefont{Press}(1999)}]{NRC}
\bibinfo{author}{\bibfnamefont{W.~H.} \bibnamefont{Press}},
  \emph{\bibinfo{title}{Numerical Recipes in C}} (\bibinfo{publisher}{Cambridge
  University Press}, \bibinfo{address}{Cambridge, U.K.}, \bibinfo{year}{1999}).

\bibitem[{\citenamefont{Cowan and Hyman}(2007)}]{c-eleg}
\bibinfo{author}{\bibfnamefont{C.~R.} \bibnamefont{Cowan}} \bibnamefont{and}
  \bibinfo{author}{\bibfnamefont{A.~A.} \bibnamefont{Hyman}},
  \bibinfo{journal}{Development} \textbf{\bibinfo{volume}{134}}
  (\bibinfo{year}{2007}).

\bibitem[{\citenamefont{Goehring et~al.}(2011)\citenamefont{Goehring, Trong,
  Bois, Chowdhury, Nicola, Hyman, and Grill}}]{Grill11}
\bibinfo{author}{\bibfnamefont{N.~W.} \bibnamefont{Goehring}},
  \bibinfo{author}{\bibfnamefont{P.~K.} \bibnamefont{Trong}},
  \bibinfo{author}{\bibfnamefont{J.~S.} \bibnamefont{Bois}},
  \bibinfo{author}{\bibfnamefont{D.}~\bibnamefont{Chowdhury}},
  \bibinfo{author}{\bibfnamefont{E.~M.} \bibnamefont{Nicola}},
  \bibinfo{author}{\bibfnamefont{A.~A.} \bibnamefont{Hyman}}, \bibnamefont{and}
  \bibinfo{author}{\bibfnamefont{S.~W.} \bibnamefont{Grill}},
  \bibinfo{journal}{Science} \textbf{\bibinfo{volume}{334}},
  \bibinfo{pages}{1137} (\bibinfo{year}{2011}).

\end{thebibliography}
\end{document}